\documentclass[aps,pra,twocolumn,superscriptaddress,amssymb,showkeys]{revtex4}

\usepackage{graphicx}

\begin{document}

%\preprint{}

\title{Phase-reversed structures in superlattice of nonlinear materials
}



\author{D.~A.~Antonosyan}
\email[]{antonosyand@ysu.am}
%\thanks{}
\affiliation{Yerevan State University, Alex Manoogian  1, 0025,
Yerevan, Armenia}\affiliation{Institute for Physical Researches,
National Academy of Sciences,\\Ashtarak-2, 0203, Ashtarak,
Armenia}

\author{G.~Yu.~Kryuchkyan}
\email[]{kryuchkyan@ysu.am}
%\thanks{}
\affiliation{Yerevan State University, Alex Manoogian  1, 0025,
Yerevan, Armenia}\affiliation{Institute for Physical Researches,
National Academy of Sciences,\\Ashtarak-2, 0203, Ashtarak,
Armenia}


\begin{abstract}
We present detailed description of so-called phase-reversed
structures that are characterized by two grating wave vectors
allowing simultaneously phase-match two parametric three-wave
processes. The novelty is that the structure is realized as a
definite assembly of nonlinear segments leading to detailed
description of cascaded three-photon processes with the parameters
of realistic structured nonlinear materials of finite length. We
apply these results for analysis of the quasi-phase-matching in
production of both photon triplet and four-photon states in
cascaded down-conversion. The received results are matched with
the experimental data.
\end{abstract}

%\pacs{42.65.Lm, 42.50.Dv, 42.65.Yj}
%insert suggested keywords - APS authors don't need to do this
\keywords{Multi-photon processes, quasi-phase-matching,
phase-reversed superlattice, down-conversion.}

\maketitle

\section{Introduction}

Cascading processes in optical materials that involve several
different second-order nonlinear interactions provide an efficient
way to lower the critical power of optical parametric devices. The
concept of multistep cascading brings new ideas into this field,
leading to the possibility of an enhanced nonlinearity-induced
phase shift, the simultaneous generation of higher-order
harmonics, multicolor parametric interaction, difference-frequency
generation, (see, \cite{Obsor} and references therein). Multistep
parametric interactions and multistep cascading are characterized
by at least two different nearly phase-matching or
quasi-phase-matching (QPM) parameters. QPM is an important
technique in nonlinear optics and is widespread used for various
applications \cite{Obsor,FejOb}. The width of the phase-matching
curve depends on the type of the crystal, its length, and the
phase-matching method. QPM not only makes efficient frequency
conversion possible, but also enables diverse applications such as
beam and pulse shaping, multi-harmonic generation, high harmonic
generation and all-optical processing. Recently, it has been
demonstrated that the quantum-computational gates, multimode
quantum interference of photons and preparation of multi-photon
entangled states also can be realized in  QPM structures
\cite{kok,Brein,RalfMun,LRMN}.

\begin{figure}
\includegraphics[width=8.6cm]{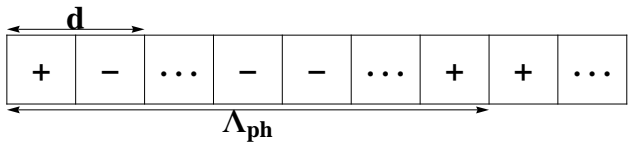}
\caption{Scheme of periodically poled nonlinear crystal (PPNC)
which involves nonlinear domains of length $l$ with $\chi_{+}>0$
and $\chi_{-}<0$ with phase-reversal grating (with  period
$\Lambda_{ph}$) upon a uniform
 QPM grating (with period $d$).} \label{revers}
\end{figure}

 One of the non-uniform QPM structures is so-called
phase-reversed structures \cite{Fejer}, which has been studied
both theoretically and experimentally and have been employed for
the wavelength conversion, and for the cascaded single-crystal
third harmonic generation process \cite{Liu,Du}. Several proposals
for increasing the phase-matching region have also been suggested,
including the use of the phase-reversed QPM structures. These
structures are based on the use of a phase-reversed sequence of
many equivalent uniform short QPM domains, particularly,
periodically poled domains, in which neighboring nonlinear
segments have the opposite signs of the second-order
susceptibility. The phase-reversed structures arranged in such a
way that at the place of the joint two ends of the domains
segments have the same sign of the $\chi^{\left(2\right)}$
coefficient (see Fig. 1). On the whole, this structure is
characterized by two grating wave vectors allowing phase-match two
processes. In the standard approach, the one-dimensional
$\chi^{\left(2\right)}(z)$ function leading to the phase-reversed
QPM structure can be described by the infinite Fourier expansion
on two periods. Such phenomenological approach qualitatively
explains the simultaneous QPM of two parametric nonlinear
processes, however, has not included the parameters of realistic
structured nonlinear materials. These dates often should be taken
into to consideration if we are interest in more detailed
description of cascaded parametric processes.

In this paper, we present more detailed description of phase-reversed structures in
another way as the definite assembly of nonlinear segments. Then,
we arrange this structure for two cascaded parametric devices
proposed for generation of three-photon and four-photon states. In
this way, we consider cascaded collinear parametric processes in
one dimensional media. The paper is arranged as follows. In Sec.
II we consider an effective nonlinear coefficient $G(\Delta k)$ of
three-wave interaction in one-dimensional phase-reversed media.
This quantity is defined as the Fourier transform of the second
order susceptibility $\chi^{\left(2\right)}(z)$ in term of wave
vectors mismatch $\Delta k$ of three-wave interaction caused by
dispersion in an optical material. In Sec.III we briefly consider
application of these results for obtaining QPM in generation of
multi-photonic states.

\section{Phase-reversed structures of finite lenght}

 We start with consideration of effective coupling constant $G(\Delta k)$ that usually
 describe three-wave interaction in second-order nonlinear media with
  the susceptibility $\chi^{(2)}(z)$ in the following form

 \begin{eqnarray}
G(\Delta k)=\int^{L}_{0}{\chi^{(2)}(z)e^{i\Delta k(z)}dz },
\label{OriginalH}
\end{eqnarray}
where $\Delta k$ is the phase-matching parameter and $L$ is a
length of medium. In the phenomenological approach the
$\chi^{\left(2\right)}(z)$ function leading to the phase-reversed
QPM structure can be described by the following Fourier expansion
\begin{equation}
\chi^{\left(2\right)}(z)=\chi_{0}\sum^{\infty}_{n=-\infty}g_{n}e^{iG_{n}z}\sum^{\infty}_{m=-\infty}g_{m}e^{iF_{m}z}
, \label{hodvaciX}
\end{equation}
that is the product of two periodic functions. Here: $g_{0}=0$,
$g_{n\neq0}=2/\pi n$, $g_{m\neq0}=2/\pi m$
    and
$G_{n}=\frac{2\pi}{d}n$, $F_{m}=\frac{2\pi}{\Lambda_{ph}}m$. In
this case the effective interaction constant reads as
\begin{widetext}
\begin{equation}
G(\Delta
k)=\chi_{0}L\sum^{\infty}_{n=-\infty}\sum^{\infty}_{m=-\infty}g_{n}g_{m}e^{-i\frac{L}{2}(\Delta
k-G_{n}-F_{m})}sinc{\left(\frac{L}{2}(\Delta
k-G_{n}-F_{m})\right)}.\label{Giranc}
\end{equation}
\end{widetext}

Considering the phase-reversed QPM structure as the definite
assembly of nonlinear segments we turn to the general formula
obtained for the effective coupling constant in multilayered media
\cite{Kly}

\begin{center}
\begin{eqnarray}
G\left(\Delta
k\right)=\sum_{m}{l_{m}\chi_{m}e^{-i\left(\varphi_{m}+\frac{\Delta{k_{m}}l_{m}}{2}\right)}sinc
\left(\frac{\Delta{k_{m}}l_{m}}{2}\right)},\nonumber\\
\varphi_{m}=\sum^{m-1}_{n}{l_{n}\Delta k_{n}},~~~~~ \varphi_{1}=0,
~~~~~~~~~~~~~~~~~~~ \label{klishkoEq}
\end{eqnarray}
\end{center}
where $\Delta k_{n}=k^{n}_{0}-k^{n}_{1}-k^{n}_{2}$ is the
phase-matching function of three photon process in the each $n$-th
domain,
$k^{n}_{j}(z,\omega)=\frac{\omega_{j}}{c}n_{n}(z,\omega_{j})$,
$n_{n}(\omega)$ is the refractive index of the medium at the given
frequency.

 In the result, the effective nonlinear
coefficient $G(\Delta k)$ for the scheme depicted in the Fig.1 can
be calculated in the following form
\begin{equation}
G \left( \Delta k\right)=L\chi_{0} e^{-i\alpha(\Delta k)}
Y_{M,N}(\Delta k). \label{P-RG}
\end{equation}
We assume that phase-matching parameters are the same for all
layers $\Delta k_{n}\equiv \Delta k$, as well as
$\chi_{+}=|\chi_{-}|\equiv \chi_{0}$.

We introduce the QPM function $Y_{M,N}(\Delta k)$ which is the
product of the standard phase-matching spectral function for a
single segment, i.e. $sinc\left(\frac{l\Delta k}{2}\right)$, and
the assemble function

 \begin{widetext}
 \begin{equation}
Y_{M,N}(\Delta k)=sinc\left(\frac{l\Delta
k}{2}\right)\frac{1}{MN}\frac{\sin{[\frac{Nl}{2}(\Delta
k-G)]}}{\sin{[\frac{l}{2}(\Delta
k-G)]}}\frac{\sin{[\frac{MNl}{2}(\Delta
k-F)]}}{\sin{[\frac{Nl}{2}(\Delta k-F)]}}. \label{YMN}
\end{equation}
\end{widetext}
  Here: $N$-is the number of domains in each block, $M$-is the
number of blocks, $\alpha(\Delta k)=\frac{Nl}{2}[(\Delta
k-G)+(M-1)(\Delta k-F)]$, $l$-is the length of each domain,
$G=2\pi  /d$, $F=2\pi /\Lambda_{ph}$, $d=2l$, $\Lambda_{ph}=2Nl$,
$L=MNl$. It is easy to check that the  function $Y_{M,N}(\Delta
k)$ has an alternative representation, as the product of
expansions on the number of domains and blocks
\begin{widetext}
\begin{equation}
Y_{M,N}(\Delta k)=sinc\left(\frac{l\Delta
k}{2}\right)\frac{1}{MN}e^{i\varphi}\sum^{N}_{n=1}{e^{-il(n\Delta
k-G_{n})}}\sum^{M}_{m=1}{e^{-iNl(m\Delta
k-F_{m})}},\label{YMNsharq}
\end{equation}
\end{widetext}
where  $\varphi=\frac{l}{2}(\Delta k-G)$. As we see, the above
expressions (\ref{P-RG}), (\ref{YMNsharq}) derived on the base of
the method of superlatices elaborated on the multilayered media
differs from the formula (\ref{Giranc}), which is obtained in the
standard, phenomenological approach. In contrast to the expression
(\ref{Giranc}), the results (\ref{YMN}),(\ref{YMNsharq}) depend on
the number of the domains and blocks and also describe the case of
small numbers of the segments. It is remarkable, that in this case
the QPM function displays specific interference effects.
Nevertheless, as shows our analysis some consequences of both
(\ref{Giranc}) and (\ref{P-RG}) results, for $N\gg 1$, $M\gg 1$,
coincide qualitatively.

Note, that single-phase-matched process of three-photon
interaction in multilayered media is also described by an assembly
function. For PPLN structure (see, one of the block in Fig.1
containing N layers) the phase-matching function is calculated as

\begin{equation}
Y_{N}(\Delta k)=sinc\left(\frac{l\Delta
k}{2}\right)\frac{1}{N}\frac{\sin{[\frac{Nl}{2}(\Delta
k-G)]}}{\sin{[\frac{l}{2}(\Delta k-G)]}}, \label{YN}
\end{equation}
where  $\Delta k$ is the wave vectors mismatch of three-wave
interaction. This function exhibits a structure of single narrow
peaks as discussed in \cite{Uren}.

\begin{figure}
\includegraphics[width=8cm]{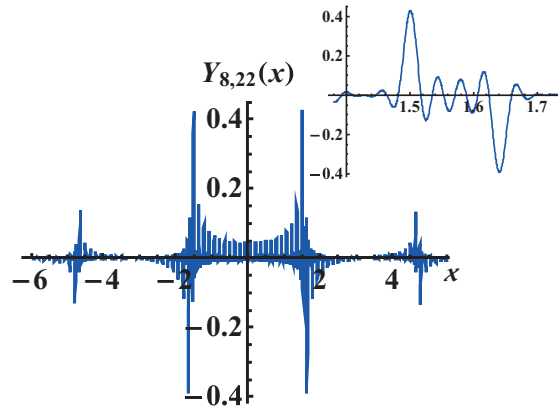}
\caption{Plot of the function $Y_{M,N}(x)$ for even numbers of
both domains and blocks; $N=22$ and $M=8$. The inset shows the
twin narrow peaks of opposite amplitudes.} \label{822}
\end{figure}

\begin{figure}
\includegraphics[width=8cm]{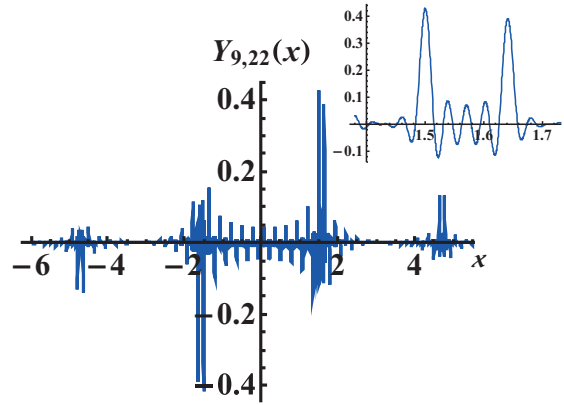}
\caption{Plot of the function $Y_{M,N}(x)$ for even numbers of
$N=22$ and odd numbers $M=9$. The inset shows the twin narrow
peaks of the same amplitudes.} \label{922}
\end{figure}

 The obtained QPM function $Y_{M,N}(\Delta k)$  allows simultaneously
phase-match two parametric processes which is achieved by using
 two grating vectors. In order to illustrate the possibility of
the double phase-matching in such superlattice structure we
present a detailed analysis of the function $Y_{M,N}(x)$ in
dependence on the parameter $x=\frac{l\Delta k}{2}$ in the
graphical form (see, Figs.~\ref{822},~\ref{922}). The assembly
function exhibits the structure of peaks that are separated by
$\pi$. Unlike to the assembly function for single-phase-matched
process, $Y_{N}(\Delta k)$ function has more complex structure,
particularly, each group of the extremes consists of thin
structure of twin-narrow peaks, separated by oscillating low
maxima. The $sinc\left(\frac{l\Delta k}{2}\right)$ multiplier
causes decreasing the amplitude of the consecutive peaks. The
behavior of $Y_{M,N}(x)$ function is different for even and odd
values of $M$. As depicted in Fig.~\ref{822} for even values of
$M$ the twin peaks have the opposite signs, while for odd values
of $M$ the both narrow peaks have the same sign (see
Fig.~\ref{922}). The existence of twin peaks and oscillation
between them are the main peculiarities of this structure. These
twin narrow peaks can be chosen to phase-match two parametric
processes involved into the multistep-cascading.


\begin{figure}
\includegraphics[width=8cm]{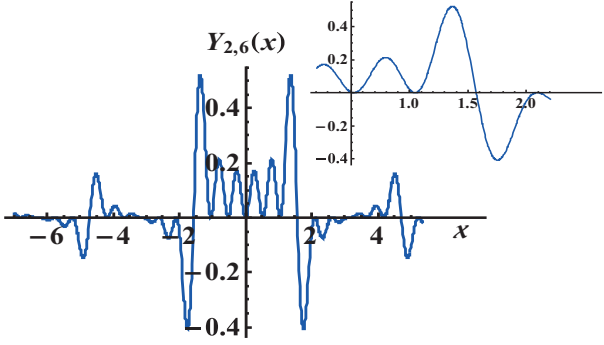}
\caption{Plot of the function $Y_{M,N}(x)$ for the small numbers
of domains and blocks, namely $N=6$ and $M=2$. The insert shows
the broadening of the peaks. } \label{Y26}
\end{figure}

 Here we represent $Y_{M,N}(\Delta k)$ function for
the small number of domains to exhibit direct correlation between
number of domains and the width of phase-matching function (see,
Fig.~\ref{Y26}). It is easy to notice that with the decrease of
number of domains and blocks the narrow peaks in the pase-matching
function becomes wider, as well as the number of oscillation
between twin peaks decreases. Besides this,  the positions of the
maxima of the phase-matching function becomes slightly shifted,
although the periodicity of the $Y_{M,N}(x)$ function is remained
the same.

\section{Applications: QPM for production of muliphoton states}

 In this section, we apply the above results
analyzing the QPM in generation of both photon triplet and
four-photon states in cascaded down-conversion. Recently,
production of joint multiphoton states, particularly, photon
triplet have attracted a great interest in probing the foundations
of quantum theory and for various applications in quantum
information technologies (see, \cite{Zeiling,Rubin,Hubel,Menq} and
the references therein). Depending on the configuration of the
experiment and the designs of the crystal, the multiphoton states
can be entangled in any number of variables: time, frequency,
direction of propagation, and polarization. Studies carried out
with miltiphoton state beams range from the examination of quantum
paradoxes, to applications in optical measurements, spectroscopy,
imaging, and quantum information (see, \cite{Rubin,TechSerg} and
references there). In this way, the direct generation of photon
triplets using cascaded photon-pairs has been demonstrated in
periodically poled lithium niobate crystals \cite{Hubel}. Most
recently, joint quantum states of three-photons with arbitrary
spectral characteristics have been studied on the base of optical
superlattises \cite{Menq}. In continuous of this paper, we analyze
the cascaded generation of photon triplets in phase-reversed
superlattice of $\chi^{\left(2\right)}$ materials.

We consider collinear, one-dimensional configurations focusing on
consideration of cascaded parametric down-conversion (PDC), which
generates three-photon states in the presence of an optical
cavity. In the scheme considered the pump field at the frequency
$\omega_{0}$ converts to the subharmonics at the central
frequencies $\omega_{1}=\frac{\omega_{0}}{3}$ and
$\omega_{2}=\frac{2\omega_{0}}{3}$ throughout two cascaded
processes: $\omega_{0}\rightarrow \omega_{1}+\omega_{2}$ and
$\omega_{2}\rightarrow \omega_{1}+\omega_{1}$. Below we consider
the amplitude of three-photon generation through the cascaded
processes assuming the frequencies $\omega_{1}$ and $\omega_{2}$
as variable quantities distributed around the central values
$\frac{\omega_{0}}{3}$ and $\frac{2\omega_{0}}{3}$, respectively.
These processes are characterized by the wave-vector mismatches:
$\Delta k_{1}=k(\omega_{0})-k(\omega_{1})-k(\omega_{2})$ and
$\Delta k_{2}=k(\omega_{2})-k(\omega^{'}_{1})-k(\omega^{"}_{1})$.
The model Hamiltonian for the system, in the rotating-wave
approximation, is given by
\begin{eqnarray}
H_{int}=i\hbar\left(E_{0}e^{-i\omega_{0}t}a^{+}_{0}-E^{*}_{0}e^{i\omega_{0}t}a_{0}\right)+\nonumber\\
i\hbar\zeta\left(a_{0}a^{+} b^{+}-a^{+}_{0}a b
\right)+i\hbar\xi\left(a^{+2} b-a^{2}b^{+}\right).\label{Hamint}
\end{eqnarray}
Here,  $\zeta$ and $\xi$ are the nonlinear coupling constants that
are the Fourier transformations of the second-order nonlinear
susceptibility $\chi^{\left(2\right)}(z)$ ( see,
Eq.(\ref{OriginalH})). The cavity modes are described by discreet
creation and annihilation operators $a^{+}, b^{+}$ and $a, b$ as
$E_{0}$ describes the amplitude of the driving fields. For the
case of the phase-reversed superlattice structure the constants of
the processes $\omega_{0}\rightarrow \omega_{1}+\omega_{2}$ and
$\omega_{2}\rightarrow \omega_{1}+\omega_{1}$ are calculated as

\begin{eqnarray}
\zeta\sim Le^{-i\alpha(\Delta k_{1})}Y_{M,N}(x_{1}),\nonumber\\
\xi\sim Le^{-i\alpha(\Delta k_{2})}Y_{M,N}(x_{2}),
\end{eqnarray}
where $x_{1}=l\Delta k_{1}/2$, $x_{2}=l\Delta k_{2}/2$.  In
general, the amplitude of three-photons at the frequencies
$\omega_{1},\omega^{'}_{1},\omega^{"}_{1}$ generated in cascaded
down-conversion has been calculated in \cite{Menq}. This quantity
is proportional to the factor $E_{0}\zeta \xi$. For the case of
phase-reverse structure we can calculate three-photon amplitude in
the following form

\begin{equation}
\Phi(\omega_{1},\omega^{'}_{1},\omega^{"}_{1})=f(\omega_{1},\omega_{2})h(\Delta
k_{1},\Delta k_{2}),\label{FipD}
\end{equation}

where
\begin{equation}
h(\Delta k_{1},\Delta
k_{2})=(MN)^2Y_{M,N}(k_{1})Y_{M,N}(k_{2})\label{h12}
\end{equation}
 is the two-dimension assemble function that describes joint phase-matching, while the function
 $f(\omega_{1},\omega_{2})=Cl^{2}\chi_{0}E_{0}(\omega_{0})e^{-i\alpha\left(\Delta
k_{1}\right)}e^{-i\alpha\left(\Delta k_{2}\right)}$, depends on
the parameters of three-wave interaction, the phase multipliers
$\alpha(\Delta k_{i})=\frac{Nl}{2}[(\Delta k_{i}-G)+(M-1)(\Delta
k_{i}-F)]$ in the formula (\ref{P-RG}) and $C$ is the
normalization factor of three-photon amplitude. Three-photon
down-conversion is controlled by energy conservation between the
pump and daughter photons
$\omega_{0}=\omega_{1}+\omega^{'}_{1}+\omega^{"}_{1}$.

Analyzing the graphics of $Y_{M,N}(x)$ (see Fig.~\ref{922}) in
application to cascaded scheme of down-conversion we could arrange
two twin peaks of the phase-matching function with the wave-vector
mismatches of cascading processes.  We demonstrate it using the
experimental dates for the cascaded down-conversion \cite{DmKZ}.
For the typical value of pump wave length $\lambda_{L}=0.53\mu m$
and LiTaO$_{3}$ crystal mismatch parameters from experimental
results are $\Delta k_{1}=0.32 \mu m^{-1}$ and $\Delta k_{2}=0.87
\mu m^{-1}$. We check that the simultaneous phase-matching of two
processes is really realized on the formula (\ref{YMN}) for the
parameters: $l=10.25 \mu m$, $N=22$, $M=9$. In this case, the
$\Delta k_{1}$ match with the second peak of the first twin-peaks
on the positive branch of the axes and the $\Delta k_{2}$ matches
with the first peak of the second twin-peaks, which are situated
next to the first twin-peaks and separated from them by the
period. We got perfect correspondence of theoretical and
experimental results.

\begin{figure}
\includegraphics[width=8.6cm]{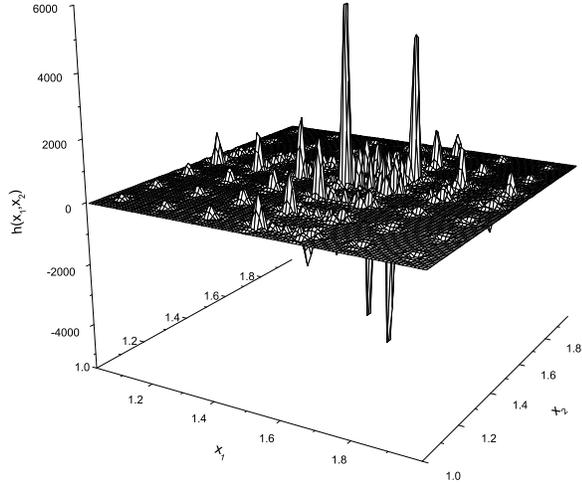}
\caption{The detailed plot of the function $h(x_{1},x_{2})$in the
range of double phase-matching peaks:  $1<x_{1}<2$ and
$1<x_{2}<2$, and the parameters $N=22$,  $M=8$. There are four
peaks: two positive , which are correspond to $x_{1}=1.5$,
$x_{2}=1.5$, $x_{1}=1.64$, $x_{2}=1.64$, and two negative at
$x_{1}=1.5$, $x_{2}=1.64$ and $x_{1}=1.64$, $x_{2}=1.5$.}
\label{1_2}
\end{figure}

\begin{figure}
\includegraphics[width=8.6cm]{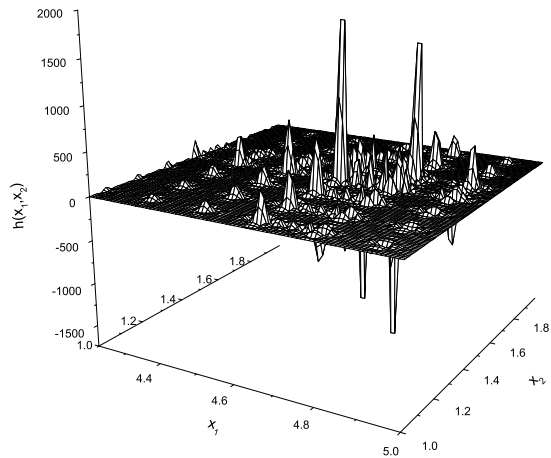}
\caption{Plot of the function $h(x_{1},x_{2})$ in the ranges
$4.2<x_{1}<5$ and $1<x_{2}<2$, and the parameters $N=22$,  $M=9$.
The heights of the peaks are three times lower than for the peaks
displayed on Fig.~\ref{1_2} in an agreement with the results
presented on Fig.~\ref{822}. } \label{4.3_1}
\end{figure}

The dependence of the spectral function $h( x_{1}, x_{2})$ on the
parameters $x_{1}$ and $x_{2}$ is shown in
Figs.~\ref{1_2},\ref{4.3_1}. As we see, the joint spectral
function displays the various maxima in $x_{1}$, $x_{2}$ place
that indicate the range of the results of the effective
three-photon splitting and perfectly confirm the results exhibited
in Fig.~\ref{822}. Thus, the above results show that the efficient
generation of three-photon subharmonic mode is possible in this
scheme. The parametric gain of the three-photon down-conversion in
the spectral range of the maxima of the function $h(x_{1},x_{2})$
is the result of the most simultaneous phase-matching of two
cascading second-order processes.

\begin{figure}
\includegraphics[width=8cm]{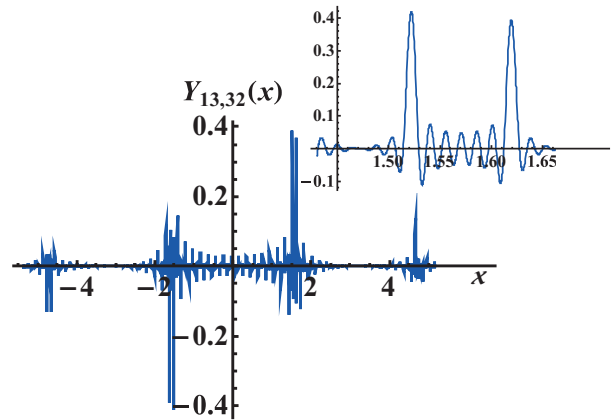}
\caption{Plot of the function $Y_{M,N}(x)$ for the parameters for
simultaneous QPM of four-photon down conversion. The number of
domains is $N=32$ and that of the blocks is $M=13$. The insert is
detailed the structure of twin-peaks. } \label{Y1332}
\end{figure}

 In the end of the section we shortly discuss QPM for generation of four-photon states in cascaded parametric oscillator. This device consists of
three intracavity modes. The pump mode driven by an external
coherent driving field at the  frequency $\varpi_{0}$ is converted
into the pair of modes $\varpi_{2}$, where
$\varpi_{0}=\varpi_{2}+\varpi_{2}=\omega_{0}/2+\omega_{0}/2$.
Then,  each of the mode $\varpi_{2}$ is transformed into second
pair of modes $\varpi_{1}$, where $\varpi_{2}=
\varpi_{1}+\varpi_{1}=\varpi_{0}/4+\varpi_{0}/4$. These processes
are characterized by the wave-vector mismatches: $\Delta
k_{1}=k(\varpi_{0})-2k(\varpi_{2})$ and $\Delta
k_{2}=k(\varpi_{2})-2k(\varpi_{1})$, respectively. To realize QPM,
we use the typical value of four-photon down-conversion pump wave
length $\lambda_{p}=0,39 \mu m$. In LiTaO$_{3}$  phase-reversed
crystal mismatch vectors are $\Delta k_{1}=1,56 \mu m^{-1}$
$\Delta k_{1}=-1,312 \mu m^{-1}$, these parameters are taken from
the experimental results for fourth-harmonic generation
\cite{Kiv4ph}. We found the parameters of the phase-reversed
structure for simultaneous realization of the four-photon
down-conversion. QPM of mentioned process is realized and
perfectly matched with the experimental results for parameters
$l=2,2 \mu m$, $M=13$ and $N=32$ (see, Fig.~\ref{Y1332}).

 \section{Conclusion}

In conclusion, we have analyzed phase-reversed QPM structures in
superlattice as the ensemble of domains and blocks of second-order
nonlinear layers. In the result, the effective nonlinear
coefficient $G(\Delta k)$ of three-wave interaction has been
obtained as the product of the standard phase-matching spectral
function for a single segment and the assemble function. Generally
used phenomenological method for description of multilayered
structures required huge number of domains and blocks to provide
the periodicity of the function. Our investigations shows that the
detailed method we have used for description of the systems
qualitatively matches with phenomenological results for the big number
of domains $N$ and blocks $M$, moreover, it is applicable for the
finite number of layers. The results shows that we can QPM various
multiphoton processes managing number of domains and blocks. We
have demonstrated that double phase-matched interaction on the
base of phase-reversed QPM structures in superlattice becomes
possible in a wide region of the optical wavelengths,
particularly, for production of muliphoton states.

\begin{acknowledgments}
This work was partially supported by ISTC, grants A-1517 and
CRDF/NFSAT/SCS No. ECSP-09-53 A-03.
\end{acknowledgments}

\end{document}